# Maxwell Matters

*Version 4, September 30, 2016*

Bob Eisenberg
Department of Molecular Biophysics and Physiology
Rush University Medical Center
Chicago IL 60612 USA

**Abstract**

Charges are everywhere because most atoms are charged. Chemical bonds are formed by electrons with their charge. Charges move and interact according to Maxwell's equations in space and in atoms where the equations of electrodynamics are embedded in Schrödinger's equation as the potential. Maxwell's equations are universal, valid inside atoms and between stars from times much shorter than those of atomic motion (0.1 femtoseconds) to years (32 mega-seconds). Maxwell's equations enforce the conservation of current. Analysis shows that the electric field can take whatever value is needed to ensure conservation of current. The properties of matter rearrange themselves to satisfy Maxwell's equations and conservation of current. Conservation of current is as universal as Maxwell's equations themselves. Yet equations of electrodynamics find little place in the literature of material physics, chemistry, or biochemistry. Kinetic models of chemistry and Markov treatments of atomic motion are ordinary differential equations in time and do not satisfy conservation of current unless modified significantly. Systems at equilibrium, without macroscopic flow, have thermal fluctuating currents that are conserved according to the Maxwell equations although the macroscopic averages of the thermal currents are zero. The macroscopic consequences of atomic scale fluctuating thermal currents are not known but are likely to be substantial because of the nonlinear interactions in systems like these, in which 'everything interacts with everything else'.

**Version 3:** Textbook models are described in Version 3, since most scientists are familiar with them, however unrealistic they are. Wording is changed a little, hopefully improved. A new section derives conservation of current from the continuity equation.

**Version 4:** Current and flux of charge in a resistor are described on p. 13.

Textbook models describe displacement currents of materials by a dielectric coefficient that is a true constant. Textbook treatments automatically satisfy conservation of current. In the textbook approximation, the displacement current of matter is proportional to $\partial \mathbf{E}/\partial t$. However, dielectric coefficients of real materials cannot be described by a constant because the displacement current of real matter is not proportional to $\partial \mathbf{E}/\partial t$ with a single proportionality constant. Even in the simplest ion solutions, dielectric coefficients vary in time by factors ~40× and displacement currents need to be described by convolution integrals with memory kernels. Dielectric properties in real matter are in fact much more complex than that: no one knows how to describe them in a general way. A general formulation of conservation of current is used here that applies whatever the properties of matter. We define 'current' as anything that creates $\mathbf{curl\ B}$. Whatever creates $\mathbf{curl\ B}$ is conserved everywhere, under all conditions, at every time. The vacuum displacement current $\varepsilon_0\, \partial \mathbf{E}/\partial t$ changes so current is conserved everywhere, under all conditions, at every time.

# Maxwell Matters

**Introduction.** Atoms have electric charge, so electricity matters.

All of life occurs in solutions with ions. Water without ions is toxic. The ions have charge. Ions are dissolved by water molecules that are strongly charged in some places, although the total charge of one water molecule is zero. The macromolecules that make the structures of life are charged. The proteins, nucleic acids (DNA and RNA), and organic molecules that are the biochemical machinery of life have electric charge, often at surprisingly high density (Jimenez-Morales, Liang, and Eisenberg 2012). Charges are likely to be highly concentrated where they are important, in technology and in life.

Charge creates forces (and moves) according to the Maxwell equations celebrated in every textbook of electrodynamics. Those equations accurately describe an enormous range of phenomena. Our electrical technology is possible because it is designed by equations, much more than by trial and error. Much of that technology depends only on Kirchoff's current law, which describes conservation of current in one dimensional systems, perhaps branched.

Kirchoff's law and electrodynamic equations can be used, without trial and error, because they are accurate and transferable from one set of conditions to another. The equations of electrical and electronic technology require little tuning of parameters as conditions change. The contrast with biochemistry is striking. The equations of biochemistry need substantial tuning as conditions change (Eisenberg 2014). The important success of semiconductor technology–that has remade our world in less than fifty years–would hardly be possible if the trial and error methods of experimental biochemistry and biophysics were used to design semiconductor devices and the integrated circuits and computers built from them.

Biochemistry (along with most of chemistry) is about making new molecules by changing their electrons and the orbitals they move in. The energy of these electrons is described by the combination of quantum mechanics and Maxwell's equations written as the Schrödinger equation. The potential term of the Schrödinger equation is a solution of the Maxwell equations.

Electrons of course are an ultimate source of charge with diameter less than $10^{-17}$ meters (Gabrielse and Hanneke 2006) and no detectable substructure. They have the remarkable property that their charge is the same, no matter how fast they move, unlike mass or spatial dimensions, all of which change as velocities approach the speed of light, according to the Lorentz transformations of special relativity. This special relativistically invariant property of electrons is confirmed experimentally in the many bright photon sources used in structural biology: those sources are synchrotrons that depend on the relativistic properties of electrons moving at speeds very close to the speed of light. (The 7 GeV beam in the Advanced Photon Source at Argonne moves electrons at >99.999999% of the speed of light. Relativistic 'corrections' are enormous $\sim 10^{16}$ at these speeds.)

But the equations of electricity do not play a prominent role in textbooks or teaching of chemistry. Maxwell equations are not found there and electricity itself is hardly mentioned in biochemistry and its textbooks, elementary or advanced.

**Maxwell Matters.** Equations (1-4) are meant to show that *Maxwell Matters* in a concise, convincing, but somewhat abstract way. This paper is a successor to a long paper arXiv:1502.0725 (Eisenberg 2016) that is more concrete than abstract, showing how *Maxwell Matters* as electricity and charge move in many situations and giving many references to the extensive literature. The discussion of Fig. 2 (Eisenberg 2016) in particular describes how charge moves in wires, salt solutions, resistors, commercial capacitors, vacuum capacitors, vacuum tube diodes, and semiconductor diodes and devices. The current through all these devices is exactly equal at all times and locations and conditions because of conservation of current (Kirchoff's law) in the series circuit, ***even though the physics of current flow is very different in each device.*** The displacement current $\varepsilon_0\, \partial \mathbf{E}/\partial t$ takes on whatever value it needs—at every time and location—to guarantee conservation of current, see eq. (4).

Maxwell introduced the idea of 'displacement current' $\varepsilon_0\, \partial \mathbf{E}/\partial t$ and the pre-existing laws of electrostatics (Coulomb's law and Poisson's equation), and magnetism (Ampere's law), could then describe the unlimited propagation of waves, including the propagation of light. Displacement current exists (and thus flows) in a vacuum in a way Maxwell found hard to visualize or explain as the movement of charge, as have many others including me.

But there is no doubt that $\varepsilon_0\, \partial \mathbf{E}/\partial t$ exists. Light exists because of the $\varepsilon_0\, \partial \mathbf{E}/\partial t$ term, so displacement current exists in a vacuum, whether or not I can imagine charges that move in the vacuum. Light exists because $\varepsilon_0\, \partial \mathbf{E}/\partial t$ helps create magnetic fields according to Maxwell's version of Ampere's law—our eq. (1)—that in turn creates electric fields. Electric and magnetic fields are coupled together to form radiation like light, that propagates through a vacuum forever.

Here I show again what has been known for a very long time: the fundamental equation of electromagnetism—Maxwell's generalization of Ampere's law eq. (1)—implies that 'current' is conserved.[1] I use a derivation and representation that emphasizes the generality of conservation of current.

**Maxwell without Matter.** Our formulation eq. (1-4) does not involve the problematic (and bewilderingly complex) properties of polarization charge and displacement current in matter. This formulation avoids the misleadingly named 'dielectric constant' $\varepsilon_r$. ***The textbook name for $\varepsilon_r$ as dielectric 'constant' seems bizarre to me*** because experiments show that $\varepsilon_r$ is almost never constant. It varies as conditions change and it always depends on time. For example, $\varepsilon_r$ varies by a factor of around 40× in sea water in experiments or in simulations starting at $10^{-15}$ sec reaching to phenomena at $10^{-6}$ sec.

The polarization charge conventionally described by a supposedly constant $\varepsilon_r$ actually varies in so many different ways that no one knows how to describe it in a reasonably general way. Formulas like eq. (2*) that involve $\varepsilon_r$ cannot be universal unless formulas for $\varepsilon_r$ are universal. Eq. (1) – (4) are universal.

**Conservation of current is universal.** Maxwell equations are true everywhere, at all times and locations, from inside atoms to between stars, we are told by physicists who spend lifetimes comparing theory and experiment.

If the Maxwell equations are universal, eq. (1-3) show that conservation of current is also universal. Conservation of current is then true everywhere, at all times and locations. ***The electric field takes on whatever value it needs—***at every time and location—***to guarantee conservation of current,*** see eq. (4). New mechanisms of current flow appear if needed to satisfy Maxwell's equations and conservation of current.

---

[1] The definition of current as whatever creates **curl B** is discussed in Section 2, along with alternative definitions.

*3*  arXiv 1607.06691 version 3 September 12 2016  *9/30/16*

The formulation of conservation of current in eq. (1-4) is important because it displays conservation in terms of universal variables without confusion from the non-universal properties of matter and $\varepsilon_r$. Eq. (2*) displays the confusion arising from $\varepsilon_r$.

The universality of conservation of current is important. Many simplified models in the literature of biochemistry and biophysics do not satisfy conservation of current (Eisenberg 2014) until they are significantly modified. Most chemical kinetic models (using the law of mass action) and Markov models using only ordinary differential equations in time suffer from this defect. All are customarily written without the displacement current term $\varepsilon_0 \, \partial \mathbf{E}/\partial t$ needed to guarantee conservation of current.

**Physical mechanisms accommodate conservation of current.** Current is conserved no matter what the detailed mechanism of that current flow.

Maxwell's equations change the value of $\varepsilon_0 \, \partial \mathbf{E}/\partial t$ (and thus the value of $\mathbf{E}$ specified by eq. 4) to force physical mechanisms to obey the conservation of current (eq. 2 & 3) exactly, no matter what the physical mechanism, or what new physical mechanism needs to appear.

The physics of charge movement is illustrated by a series circuit made of wires, salt solutions, resistors, commercial capacitors, vacuum capacitors, vacuum tube diodes, and semiconductor diodes shown in Fig. 2 of (Eisenberg 2016). The mechanism of current flow is very different in each component and varies with time in dramatic ways in some of them. These diverse mechanisms have the same current through them at any time, as is necessary if they are to follow Maxwell's equations. It is a miracle that these distinct mechanisms can each change in its own way so current is conserved.

Continuity of current does not arise as a consequence of atomic properties or of conservation of matter. It arises because of the universal properties of electrodynamics.

Conservation of current is exact because of the $\varepsilon_0 \, \partial \mathbf{E}/\partial t$ term in Maxwell's equation (1). The $\varepsilon_0 \, \partial \mathbf{E}/\partial t$ does not involve material properties. The $\varepsilon_0 \, \partial \mathbf{E}/\partial t$ term can vary so that Maxwell's equations and current conservation are exact, at every time and location. The time rate of change of the electric field $\partial \mathbf{E}/\partial t$ can take on whatever value it needs (at any time and any location) to guarantee that current is conserved exactly everywhere and at every time.

Amazingly, the abstract conservation law (eq. 2) changes the atomic properties of systems and sometimes even introduces new physics and mechanisms into a system. In every day experience, $\partial \mathbf{E}/\partial t$ and electric fields can become large enough to change the physics of charge flow, introducing phenomena not in the original intuitive description of circuits.

Pulling the power plug from the electrical socket in a wall can create electric fields large enough to create a spark. The fields ionize air, creating a plasma that sparks in the dark. The electric fields become large enough to destroy the insulating properties of air. *Equations of the normally insulating air must be changed if we wish to describe the spark we observe*. *The constitutive equations need to be generalized to deal with the effects of the enormous electric field needed to preserve continuity of current.* Maxwell's equations create (seemingly) new physics.

**Maxwell Matters in the models and theories of science** in a way not widely recognized. *Current is conserved everywhere, so it must be conserved in the theories and models of science.* Many of the models and theories of science are simplified so they depend only on time, using ordinary differential equations in time. These models should satisfy conservation of current. They do not satisfy conservation of current unless additional constraints are added to their usual formulation.

**Kinetic models: mass action and Markov.** Sadly, the equations of chemical and enzyme kinetics, based on the 'law' of mass action, and the equations of Markov models in general, are ordinary differential equations in time that do not satisfy conservation of current unless modified.

A simple computation of the charge that moves in kinetic or Markov scheme $\mathbf{A} \rightleftarrows \mathbf{B} \rightleftarrows \mathbf{C}$ (Eisenberg 2014) shows that the current from **A** to **B** does not equal the current flowing from **B** to **C** unless additional constraints are imposed. The additional constraints are likely to change the *qualitative* nature of the kinetic or Markov scheme because it needs to include the displacement current $\varepsilon_0\, \partial \mathbf{E}/\partial t$ and needs to include spatial dependence with its boundary conditions.

**Models in textbooks of electrodynamics** have inadvertently obscured the general significance of conservation of current in my opinion. Textbooks present Poisson's equation with a dielectric coefficient $\varepsilon_r$ (dimensionless, ~79 for water) and permittivity of matter that is a constant, really a constant, a single number. In that case, it is easy to apply the continuity equation for the flux of charge and show that material displacement current is $\varepsilon_r \varepsilon_0\, \partial \mathbf{E}/\partial t$. No distinction between material displacement current and vacuum displacement current is needed.

The textbook treatment is unfortunately a serious distortion of reality. The dielectric coefficient is never constant independent of time. In biological and chemical applications it varies by a factor of 40× in the time scales of interest (linking atomic motions at $10^{-16}$ sec and biological function starting at $10^{-4}$ sec). And the properties (even the meaning) of dielectric coefficient/permittivity is very different in different systems in which current flows. The permittivity of a dielectric capacitor and a vacuum capacitor have little to do with each other; the permittivity of the various materials described in Fig. 2 of arXiv:1502.0725 (Eisenberg 2016) are all variables not constants and each varies in quite different ways. A general description of permittivity and dielectric coefficient of all the materials in a typical electrical circuit is not easy to find, if it is in fact known, or possible at all.

The general role of conservation of current is obscured by models that pretend matter can be described by a constant dielectric coefficient. Only when Maxwell's equations are written as in eq. (1) without invoking the properties of matter, does the significance of conservation of current become clear.

**History.** Kinetic schemes of chemical reactions, and of Markov models use ordinary differential equations that depend only on time. They ignore differential terms in space. The simplification to deal with just time is more than understandable historically. Ordinary equations could be understood and computed long ago, before computers and numerical methods were available to deal with partial differential equations in space and time. But the laws of physics do not change because of our technological and intellectual limitations. *Conservation of current involves space. Models and theories that use only time derivatives need to conserve current.*

**Equilibrium systems.** Maxwell matters even in equilibrium thermodynamic systems that seem at first not to involve current flow, because all systems, including thermodynamic systems at equilibrium without macroscopic flow, are made of atoms in thermal motion. These thermal motions produce local currents even in systems without global current flow. The thermal motion of the charged atoms that make up matter produce local variations in the density of charge and thus in the local electric field. These local variations are different in different places. (After all, that is what the word 'local' means.) Spatial variations in electric field imply current flow. Molecular dynamics simulations visualize these current flows and confirm their size and importance. Roughly speaking, simulations of ionic solutions show changes of electrical potential of ~1 volt in intervals much shorter than those involved in most of biology or chemical technology, in intervals of a few picoseconds ($10^{-12}$ sec). Thermal currents flow on similar scales.

The ***fluctuating thermal currents are conserved*** just as all current is conserved. Even though the thermal fluctuations average to zero, they need to satisfy equations (1)-(2) ***in every time interval and region of space, no matter how brief or small.***

Thermal fluctuations are described by coupled nonlinear field equations in which 'everything interacts with everything else.' The thermal fluctuations in current are likely to have effects on many time and length scales as the nonlinearity converts atomic fluctuations into macroscopic phenomena.

The significance of these thermal fluctuating currents in thermodynamic equilibrium systems—in systems without macroscopic flow—is not clear to me. In some cases the fluctuating thermal currents will average out and not have macroscopic effects (although atomic scale effects will be observable by experiments with atomic resolution). But the question remains what are those cases? What are the situations in which the fluctuating thermal currents do not average out? The implications of conservation of current in thermodynamics remains to be investigated in general. Dave Ferry has pointed out a specific case where the thermal fluctuating currents have steady state consequences (Ferry 1980).

# Conclusion: Maxwell Matters

**Maxwell's equations need to be included in all our models** and theories, in my view. Approximate and reduced models need to respect conservation of current with little error because of the enormous strength of the electric field in general, so vividly described in the third paragraph of Feynman's textbook on electricity and magnetism (Feynman, Leighton, and Sands 1963). Specific estimates of the strength of the electric field are described in (Eisenberg 2016).

## Section 1: <u>Conservation of Current</u>[2]: derivation from Ampere's law

Maxwell's generalization of Ampere's law ensures the existence of (a wave equation for) light. Note that the variable **J** includes all movements of charge. It includes what is usually called polarization current as well as transport current.

> One could subtract ideal polarization current $(\varepsilon_r - 1)\varepsilon_0 \partial \mathbf{E}/\partial t$ if one wished (with $\varepsilon_r$ a constant: see next section of this paper) but the polarization of real materials is not well approximated by that term and so $\mathbf{J} - (\varepsilon_r - 1)\varepsilon_0 \partial \mathbf{E}/\partial t$ would contain a description of the nonideal properties of polarization in real materials. I prefer not to separate the $(\varepsilon_r - 1)\varepsilon_0 \partial \mathbf{E}/\partial t$ term since it does not describe experimental reality.

$$\mathbf{curl}\left(\mathbf{B}/\mu_0\right) = \underbrace{\mathbf{J} + \varepsilon_0 \frac{\partial \mathbf{E}}{\partial t}}_{\text{'Current'}} \qquad \text{\textit{Note the subscript zero on }} \varepsilon_0. \text{ \textit{It is crucial.}} \qquad (1)$$

In math language,

$$\mathbf{div}\underbrace{\left(\mathbf{J} + \varepsilon_0\, \partial \mathbf{E}/\partial t\right)}_{\text{'Current'}} = 0 \qquad \text{\textit{because \textbf{div curl} = 0 is an identity.}} \qquad (2)$$

In physical language,

$$\boxed{\underbrace{\mathbf{J} + \varepsilon_0 \frac{\partial \mathbf{E}}{\partial t}}_{\text{'Current'}} \text{ is conserved exactly, always, everywhere.}} \qquad (3)$$

In American/English

> Anything that creates **curl B** is current.
> Anything that creates **curl B** is conserved.
> Current creates **curl B**. Current is conserved.

Eq. (1) provides an operational definition of **J** that **allows J** to be defined by experiments wherever **curl B** can be measured. Eq. (2) shows that conservation of current can be written without reference to the properties of matter.

**Remark 1:** The polarization current $(\varepsilon_r - 1)\varepsilon_0 \partial \mathbf{E}/\partial t$ of an hypothetical ideal polarizable material (with $\varepsilon_r$ constant) can be separated and subtracted from **J** but the polarization of real materials is not well approximated by the ideal expression. The difference between real and ideal is $\mathbf{J} - (\varepsilon_r - 1)\varepsilon_0 \partial \mathbf{E}/\partial t$. The difference describes the **non**ideal properties of polarization of actual materials. The nonideal polarization is much larger than the ideal polarization, roughly corresponding to the ratio of $\varepsilon_r$ on atomic and macroscopic time scales. That ratio is $\cong 40\times$ in ionic solutions. I prefer not to separate the ideal term $(\varepsilon_r - 1)\varepsilon_0 \partial \mathbf{E}/\partial t$, with $\varepsilon_r$ constant, since the ideal term does not describe experimental reality and so does not deserve special treatment by isolation.

---

[2] The definition of current is discussed in Section 2, along with alternative definitions.

I prefer to describe **all** the polarization properties of a material (both ideal and nonideal, the sum being the total real polarization properties measured in experimental impedance/admittance measurements) in just one component which is part of the **J** defined in eq. (1) and used in eq. (2).

The total actual polarization of each material (as measured in experiments) then is described by its own component that describes all movement of charge, whatever its dynamics, whether they depend on $\partial \mathbf{E}/\partial t$ in part or not.

The **part** of the total actual polarization of each material (as measured in experiments) that is strictly proportional to $\partial \mathbf{E}/\partial t$ (with a single proportionality constant that is a single real number) then corresponds to just the ideal part of polarization, namely the component $(\varepsilon_r - 1)\varepsilon_0 \partial \mathbf{E}/\partial t$ (for that material). That ideal component (but only that ideal component) can be described by a constant dielectric coefficient $\varepsilon_r$ where the constant is a single real number. That ideal component is a small part of the total polarization of most materials and so (to repeat) does not deserve isolation and special treatment in the theory, in my view.

**Remark 2:** The special term $\varepsilon_0 \partial \mathbf{E}/\partial t$ describes what happens in empty space, where $\mathbf{J} = 0$. $\varepsilon_0 \partial \mathbf{E}/\partial t$ is sometimes called the vacuum displacement current. In a vacuum, no matter interferes with the perfect time-independent linear relation between 'current' and the magnetic and electromagnetic field. The 'current' is a linear function of **E** that is instantaneous (without delay) with a proportionality constant that is truly constant, independent of anything.

*The special term $\varepsilon_0 \partial \mathbf{E}/\partial t$ also exists inside matter:* atoms are almost all empty space (with their mass concentrated into a tiny nucleus). Maxwell equations are 'universal' in vacuum and in matter, from inside atoms to between stars, from time scales of $\gamma$-rays (even cosmic rays) to steady state, with an accuracy of some 1 part in $10^{18}$.

## Summary

Equation (2) says

---

'Current' defined as $\mathbf{J} + \varepsilon_0 \partial \mathbf{E}/\partial t$ is conserved exactly, always, everywhere

**E** can be whatever it needs to be to ensure that $\mathbf{div}(\mathbf{J} + \varepsilon_0 \partial \mathbf{E}/\partial t) = 0$.

**E** can be whatever it needs to be because of the vacuum term $\varepsilon_0 \partial \mathbf{E}/\partial t$.

The vacuum term $\varepsilon_0 \partial \mathbf{E}/\partial t$ is *independent of matter.*

so $\partial \mathbf{E}/\partial t$ can assume *any value it must* to ensure $\mathbf{div}(\mathbf{J} + \varepsilon_0 \partial \mathbf{E}/\partial t) = 0$, no matter what are the properties of matter and **J**.

---

## Section 2: **Conservation of Current:** Derivation from Continuity Equation

Conservation of current is often deduced from the continuity equation, as in the classical text of Jackson (Jackson 1999), in eq. 6.4 on p. 238.

$$\mathbf{div}\,\mathbf{J}_{matter} + \frac{\partial \rho}{\partial t} = \mathbf{div}\left(\mathbf{J}_{matter} + \frac{\partial \mathbf{D}}{\partial t}\right) = 0 \tag{1*}$$

[Jackson's symbol for flux is replaced by $\mathbf{J}_{matter}$ to avoid confusion with $\mathbf{J}$ in eq. (1).]

$$\mathbf{D} \text{ is defined as } \mathbf{D} = \varepsilon_0 \mathbf{E} + \mathbf{P}, \text{ giving} \tag{2*}$$

$$\mathbf{div}\left(\mathbf{J}_{matter} + \frac{\partial(\varepsilon_0 \mathbf{E} + \mathbf{P})}{\partial t}\right) = \mathbf{div}\Bigg(\underbrace{\underbrace{\mathbf{J}_{matter} + \frac{\partial \mathbf{P}}{\partial t}}_{\mathbf{J}} + \varepsilon_0 \frac{\partial \mathbf{E}}{\partial t}}_{\text{'current'}}\Bigg); \quad \mathbf{J} \text{ is defined in eq. (1)} \tag{3*}$$

The conserved quantity 'current' is then

$$\text{'current'} = \mathbf{J}_{matter} + \frac{\partial \mathbf{P}}{\partial t} + \varepsilon_0 \frac{\partial \mathbf{E}}{\partial t} \tag{4*}$$

**Idealization implies consistency.** The polarization term $\partial \mathbf{P}/\partial t$ is often written as if materials were ideal. If materials were ideal, we could write $\partial \mathbf{P}/\partial t$ as $(\varepsilon_r - 1)\varepsilon_0 \partial \mathbf{E}/\partial t$ with a constant dielectric coefficient $\varepsilon_r$.

The idealization has an important advantage beyond simplicity. If $\varepsilon_r$ is a constant, the electrostatic equation $\mathbf{div}\,\mathbf{D} = \rho$ becomes $\mathbf{div}\,\mathbf{E} = \varepsilon_r \varepsilon_0 \rho$ and automatically satisfies conservation of current as we can see by differentiating with respect to time and combining with the continuity equation (1*) using $\partial \mathbf{P}/\partial t = (\varepsilon_r - 1)\varepsilon_0 \partial \mathbf{E}/\partial t$.

The idealized model is fully consistent. Every variable satisfies every equation and boundary condition, under all circumstances, with one set of parameters. Very large errors quickly arising from inconsistent treatments of charge and current are avoided (Appendix of Eisenberg, 2016). Errors are found in treatments of flow that do not include the vacuum displacement current $\varepsilon_0 \partial \mathbf{E}/\partial t$. Such treatments are not rare since many scientists identify electric current with the flux of charges, thereby ignoring the vacuum displacement current $\varepsilon_0 \partial \mathbf{E}/\partial t$ needed to guarantee consistency and the universal conservation of current. (An example is given in Eisenberg, 2016, p. 10, **Resistor** section and discussed later in this paper, on p. 13).

**Over-simplification** is a necessary part of science. The over-simplification that $\varepsilon_r$ is a constant is a fib—a 'white lie'—that is useful in teaching because it allows so much to be done with minimal mathematics. And the over-simplification has the important advantage of guaranteeing consistency. But statements of conservation of current that include $\varepsilon_r$ as a constant are drastic over-simplifications if they do not include $\mathbf{P}_{excess}$ (see eq. 5*) or its equivalent. Materials are not ideal, not even approximately in the time/frequency domain of our technology ($10^{-9}$sec), let alone of our atomic scale simulations of molecular dynamics ($10^{-15}$sec). Polarization and dielectric properties vary by factors of 40× in those ranges, and delays and overshoots in charge movements are prominent, so alternative formulations are required.

Statements of conservation of current that include $(\varepsilon_r - 1)\varepsilon_0$ should be avoided (in my opinion) because they are only valid for the tiny range of systems and conditions in which $(\varepsilon_r - 1)\varepsilon_0$ is constant.

**Consequences of Over-simplifications.** Over-simplifications are a necessary part of science, as we have said. But they are two edged swords. They simplify and guarantee consistency. But fibs undermine credibility. The over-simplification of $\varepsilon_r$ as a constant hides the universal nature of conservation of current (see Appendix).

'Everyone' knows that $\varepsilon_r$ is not constant, so 'everyone' thinks that equations involving $\varepsilon_r$ are over-simplifications and not universal laws at all. 'Everyone' then thinks that conservation of current is an over-simplification, and not a universal law.

But conservation of current in not an over-simplification. Conservation of current is a universal law. It can be written without reference to any property of matter, or any constitutive law, as in eq. (3). Indeed, eq. (4) shows that the electric field takes on whatever value it needs to ensure conservation of current, independent of any particular constitutive equation. The electric field will change that constitutive equation, or enlist a 'new' phenomena (e.g., ionization of air) with its own constitutive law, if it is needed to ensure conservation of current.

**Polarization.** In the classical literature of polarization (using analysis of sinusoids in the frequency domain), the dielectric constant $\varepsilon_r$ (a real number) is often generalized into a complex number with amplitude and phase (or real and imaginary parts). Polarization of this type is described in the time domain by solutions of ordinary differential equations (usually with constant coefficients) or convolution integrals (with memory kernels). (The convolution integrals arise from inverse Fourier-Laplace transforms of the frequency dependent $\varepsilon_r$.)

In my view, it is much more useful to **define** the dielectric coefficient $\varepsilon_r$ **always** as a real number equal to its value at DC (very low frequencies or long times) and write current as

$$\text{'current'} = \mathbf{J}_{matter} + \underbrace{\frac{\partial \mathbf{P}_{excess}}{\partial t} + (\varepsilon_r - 1)\varepsilon_0 \frac{\partial \mathbf{E}}{\partial t}}_{\text{Material Displacement}} + \varepsilon_0 \frac{\partial \mathbf{E}}{\partial t}; \quad \varepsilon_r \text{ a real constant} \qquad (5^*)$$

The complex properties of matter are then dealt with in the two terms $\mathbf{J}_{matter} + \partial \mathbf{P}_{excess}/\partial t$.

Or we can dispense with the polarization term altogether and include it in the flux of matter $\mathbf{J}$. Then we have eq. (1)–(2),

$$\text{'current'} = \mathbf{J} + \varepsilon_0 \frac{\partial \mathbf{E}}{\partial t} \qquad (6^*) \text{ or } (3)$$

albeit with a different definition of $\mathbf{J}$ from that in classical texts (Jackson 1999).

**Polarization flux is more complex than it seems.** Polarization flux includes subtle and complex correlations and coupled transport of considerable importance.

Dielectrophoresis is an example of complexities that occur in real ionic systems. Consider dielectrophoresis of 'uncharged' particles in an ionic solution. In dielectrophoresis, particles with zero permanent charge can be transported by the electric field because the particles have induced polarization charge. That is to say, in formal terms, $\partial^2 \mathbf{E}/\partial x^2 \neq 0 \Rightarrow \mathbf{J}_{\text{transport = dielectrophoresis}} \neq 0$. Phenomena like dielectrophoresis produce **both** transport $\mathbf{J}_{\text{matter}}$ and material displacement current defined in eq. (5*) as $\partial \mathbf{P}_{\text{excess}}/\partial t + (\varepsilon_r - 1)\varepsilon_0\, \partial \mathbf{E}/\partial t$.

I prefer the formulation eq. (2) to that of eq. (5*) because eq. (2) does not separate one physical phenomenon (like dielectrophoresis) into misleadingly distinct terms described by two apparently independent constitutive equations, one for $\mathbf{J}_{\text{matter = dielectrophoresis}}$; the other for $\varepsilon_r - 1$.

**Variational methods.** In my opinion, coupled phenomena like dielectrophoresis need to be described by energy variational methods that derive field equations from an Euler Lagrange process (Ryham, Liu, and Wang 2006; Ryham 2006; Eisenberg, Hyon, and Liu 2010; Horng et al. 2012; Forster 2013; Wu, Lin, and Liu 2014b, 2014a; Xu, Sheng, and Liu 2014; Wu, Lin, and Liu 2015; Wang, Liu, and Tan 2016). The Euler Lagrange process ensures that all cross terms in constitutive equations are consistent.[3]

Attempts to separate transport and material terms before deriving the Euler Lagrange field equations are tricky and likely to lead to error. Cross terms are easy to overlook, inadvertently. The definition of 'transport' and 'material displacement' is unlikely to be unique before Euler Lagrange equations are actually derived from a variational principle. Different workers are likely to classify components of transport and displacement in different ways until the Euler Lagrange is used to display explicitly all the components needed in a self-consistent analysis.

*After* the variational analysis is completed and Euler Lagrange equations have been actually derived, coupled phenomena like dielectrophoresis are described by a unique set of field equations and boundary conditions, including cross terms. Coupled phenomena can **then** be separated into transport and material displacement terms and workers can agree on the names and classification of cross terms.

**Charge movement in general.** The general case requires a general treatment of **J** because it involves charge movement much more diverse and complicated than those of polarization phenomena.

A complete description of charge movement is different—very different—in different systems, because the properties of charge are as diverse as the properties of matter. (After all, matter is made of charges and any field induced movement of charge by any type of field must be included in the equations, e.g. eq. 1*–4*.) If charge movement is coupled to other fields (e.g., diffusion, flow or temperature), an even more complex treatment is needed using variational methods (in my opinion).

Even the (apparently) simplest consistent treatment of the diffusion and migration of charges (Poisson Nernst Planck, called Poisson drift diffusion in the semiconductor literature) involves the complex mathematics of field theory. PNP is nonlinear, time dependent, and couples 'everything to everything else'. The range of behavior that PNP can describe is astonishing. PNP can produce all the nonlinear functions found in digital logic and the analog functions needed to use them. The spatial distribution of permanent charge (created by doping) can be designed to make semiconductor devices with robust properties. PNP then produces the wide range of semiconductor devices found in analog and digital computers and all the nonlinear functions found in their digital logic. The properties of charge movement described by PNP are in fact diverse. They are diverse enough to describe anything a computer can store or do.

---

[3] 'Consistent' means that all independent variables satisfy all field equations and boundary conditions in all circumstances. In such systems electrical potential depends on charges. Potential profiles and rate constants that are constant, independent of conditions, are found in inconsistent models like the constant field or rate constant models of biophysics. Potential profiles and rate constants of those models do not change when charge or structure changes.

No general description of charge movement for all matter is known, or, in my opinion, likely to be discovered. Matter just comes in too many forms and its charge movement are just too diverse for that.

## Section 3: Formula for E that Ensures Conservation of Current

We now write a formula for the electric field that guarantees that Maxwell eq. (1) and conservation of current (2) are satisfied. We start with equation (2) restated here

$$\mathbf{div}\left(\mathbf{J} + \varepsilon_0\, \partial \mathbf{E}/\partial t\right) = 0 \tag{2}$$

Eq. 2 is satisfied if

$$\mathbf{E} = -\int \left(\mathbf{J}/\varepsilon_0\right) dt, \tag{4}$$

In words, if the electric field is the time integral of the flow, conservation of current is satisfied. The general solution of eq. (2) is much more involved, as Carl Gardner kindly pointed out. This particular solution (4) is always available to the system, so eq. (4) is enough to guarantee that **E** can take on whatever value is needed to produce conservation of current.

In summary

> **If the electric field in a vacuum**,
> including the vacuum within atoms is
>
> $$\mathbf{E} = -\int \frac{\mathbf{J}}{\varepsilon_0} dt, \tag{4}$$
>
> the conservation of current defined in eq. (2) is exact,
> **no matter what the constitutive equation for J**.

# Section 4: Kirchoff's Current Law

Kirchoff's current law is a restatement of eq. (2) appropriate for electrical circuits. Electrical circuits are one dimensional systems that are usually branched, but hardly ever extend significantly into three dimensions. They do not have the spatial singularities characteristic of Maxwell's equations in two or three dimensions even when they include transmission lines described by the telegrapher's (partial differential) equation. Most of our electrical, electronic, semiconductor, and computer technology is performed by one dimensional branched circuits that can be described by Kirchoff's current law, Kirchoff's voltage law, and nothing else.

In unbranched one dimensional systems of components (i.e., in series circuits) the 'current' of our eq. (2) is the current of Kirchoff's current law and is independent of location. In the series circuit of Fig. 2 (Eisenberg 2016), the current is the same at every time and every location in the wires, salt solutions, resistors, commercial capacitors, vacuum capacitors, vacuum tube diodes, and semiconductor diodes and devices although the physics of each of these devices is very different. The electric field changes the specific physics of each device so the current is exactly equal. Indeed, the electric field will recruit new physical phenomena if that is the only way to satisfy Maxwell's equation, as it does when a power plug is pulled out of the electrical socket in the wall, when the **E** field becomes strong enough to strip electrons from the atoms of air, making an insulator into a plasma.

**Ohm's law and Kirchoff's law do not accumulate charge.** Circuits containing only ideal resistors do not allow charge storage, as discussed in Eisenberg, 2016 (p. 10, **Resistor** section).

The flux of charges (without the displacement term $\varepsilon_r \partial \mathbf{E}/\partial t$ or $\varepsilon_0 \partial \mathbf{E}/\partial t$) that flow into a resistor equals the flux of charges (without the displacement term) leaving the resistor at all times, and at every time, so there is no difference between input and output flux of charges (without the displacement term) and no build-up of net charge. In electrostatics only net charge can produce potential, according to Maxwell's electrostatic equation—e.g., Poisson's equation—so a one dimensional series circuit has no net charge anywhere and no potential anywhere if it is made only of resistors that strictly obey Kirchoff's current law (at all times) for flux of charges (without the displacement term).

Ohm's law (that defines a resistor in this discussion), and Kirchoff's current law (applied to just the flux of charges without the displacement term $\varepsilon_r \partial \mathbf{E}/\partial t$ or $\varepsilon_0 \partial \mathbf{E}/\partial t$) do not provide the net charge needed to create a potential (in Maxwell's electrostatic equation).

**Kirchoff's law can be generalized.** Net charge appears in one dimensional series circuits if Kirchoff's current law is generalized to describe current that includes the displacement term $\varepsilon_r \partial \mathbf{E}/\partial t$ or $\varepsilon_0 \partial \mathbf{E}/\partial t$ instead of flux of charges without the displacement term. When current (including the displacement term) is used as the variable in Kirchoff's law, charge builds and potentials change.

**Capacitors can be added.** Alternatively, one can use the classical form of Kirchoff's law (that describes the flux of charges without the displacement term $\varepsilon_r \partial \mathbf{E}/\partial t$ or $\varepsilon_0 \partial \mathbf{E}/\partial t$) by adding capacitors (somewhat artificially) to the circuit that originally contained only resistors. Potentials are then produced by charge on the capacitors as required by Poisson's equation. The capacitors introduced at each node allow current $i_c$ to flow from node to node or node to ground with $i_c = C \partial V/\partial t$ where $V$ is the electrical potential and $C$ is a true constant perhaps different for each capacitor. These capacitors introduce a variable potential $V$ that allows the branched one dimensional system to be consistent with *V* computed from Maxwell's equations, including conservation of current and charge. Branched circuits are studied with Kirchoff's current law using the techniques of circuit theory.

**Adding capacitors to kinetic and Markov models.** This approach can be applied to classical models that do not include the displacement term $\varepsilon_r \, \partial \mathbf{E}/\partial t$ or $\varepsilon_0 \, \partial \mathbf{E}/\partial t$ so the classical models can be modified to satisfy conservation of current. For example, rate models (of chemical or Markov theories) can be modified by adding capacitance into each node of the rate model. The capacitance connects the node to 'ground' and introduces what the chemists misleadingly call a self-energy, even though that energy depends on the location and strength of a dielectric boundary that is not a property of 'self' at all.

The details of the capacitor representations need to be worked out to see if they are worth the increase in complexity of the resulting expressions for flux and current. It is not clear that introducing capacitors will be easier than just solving the 'real' problem involving partial differential equations in space and time.

The revised representations will be rather involved because the expressions for rate constants often must include the shape of potential barriers, as well as the height of the barriers. The flow over such barriers can be computed with simple integrals as shown in Eisenberg, 2011, but the integrals depend on the shape of the potential barrier and that is not a local variable. The shape of the potential barrier depends on boundary conditions and on more than one 'reaction' in the rate scheme so spatial dependence is introduced into the system and problem, as it must be if current flows. Considerable complexity is expected.

# Section 5: Equilibrium Systems without Current Flow
Equilibrium Systems

The discussion of the previous pages seems to apply only when current flows. It might seem not to apply to classical equilibrium systems of thermodynamics or to systems in which the flow of current is zero.

I argue here, however, that conservation of $\mathbf{J} + \varepsilon_0 \, \partial \mathbf{E}/\partial t$ is important in systems without net flow, even in classical equilibrium systems of thermodynamics because ***all systems have atomic scale flows of current arising from thermal motion.*** These flows need to satisfy the Maxwell equations because those are universal. These facts are obscured by the many classical treatments of Brownian motion in which electric fields are assumed to be independent of time, location, or charge density.

In my opinion, Brownian motion needs to be computed self-consistently, so the electric field varies with charge density, and thus with time and location, and electromagnetic phenomena are computed from the Maxwell equations (Eisenberg 2006). In this proper treatment of Brownian motion, thermal fluctuations in the density of charge must produce thermal fluctuations in the electric field. Fluctuations in the electric field imply that $\partial \mathbf{E}/\partial t \neq 0$. They must produce current according to eq. (1)-(3). Those fluctuating thermal currents $\mathbf{J} + \varepsilon_0 \, \partial \mathbf{E}/\partial t$ are conserved at every point in time and space by eq. (2-3), just as all currents are conserved. The stochastic thermal currents are local and fluctuate around a zero mean, since the space and time averages clearly have to be zero in thermodynamic systems, or systems with zero macroscopic $\mathbf{J} + \varepsilon_0 \, \partial \mathbf{E}/\partial t$. But the combination of thermal motion and Maxwell equations support very nonlinear phenomena and it is clear that the nonlinear phenomena can extend to macroscopic scales (Ferry 1980).

**Chemical reactions do not occur uniformly in ionic solutions.** It is important to see the consequences of thermal fluctuating currents in a thought experiment of a chemical reaction in an ionic solution.

Imagine a chemical reaction that depends steeply on concentration of reactants (as many do, but certainly not all). Many reactions occur on a time scale of $10^{-16}$ or faster because they involve movements of electrons in molecular orbitals that are not delayed by inertia or friction. Clearly, these reactions will occur mostly in the regions of the solution that have large concentrations of reactants. Thermal fluctuations in concentration occur on time scales of say $10^{-9}$ seconds over regions involving large numbers of atoms. Regions of large concentration exist for times very much longer than the time of the chemical reaction itself. There is plenty of time for chemical reactions to occur in regions that thermally fluctuate to large concentrations. It is clear that *the concentrations of the molecules actually reacting will not be the spatially average concentration.* Chemical reactions in ionic solutions will occur mostly in regions that have concentrations far from the spatial average.

We expect that chemical reactions occur nonuniformly in ionic solution. The reactions are different in different locations. They occur mostly in special places where concentrations have fluctuated to large values. Theories that only use average values of concentration will not describe what is happening in these situations. Theories are most useful when they are robust and can be transferred from one situation to another without changing parameters. Theories of chemical reactions in solutions need to include spatial and temporal thermal variations, and concomitant thermal fluctuating flows of current (eq. 1-3). Everything is coupled to everything else in such systems and they are usually very nonlinear because of the coupling. Theories need to describe that coupling and the resulting nonlinearity. Only energy variational methods (that include dissipation) can do that, as far as I know.

**Kinetic Models of Biochemical Reactions** rarely fit experimental data if conditions are changed, even in quite simple ways (Eisenberg 2014). Models are said in chemical language to be not 'transferable' if different rate constants are observed in different situations.

Rate constants of chemical reactions observed experimentally usually change with (1) changes in 'background' salt concentration—by adding $Na^+Cl^-$ for example; (2) changes in the type of 'background' ion (from say $K^+Cl^-$ to $Na^+Cl^-$); (3) changes in divalent concentration (typically $Ca^{2+}$ or $Mg^{2+}$); even with (4) changes in concentrations of the reactant or enzyme.

Rate constants imply free energies. Changes in rate constants imply changes in free energies in the underlying chemical reaction. When changes in rate constants are needed to fit data in a new set of conditions, the original chemical reaction and its kinetic model need an additional free energy to fit experimental data in those new conditions. The last sentence is a polite way of saying the original reaction scheme is incomplete, and in that sense incorrect.

Reaction schemes are not transferable because they leave out a component of free energy. Rate constants change as the reaction is transferred from one set of conditions to another because the reaction scheme and kinetic model leave out a component of free energy. The new model needs to adjust its rate constants to make up for the free energy it has left out.

Kinetic models can be revised so they are consistent with Maxwell equations. Perhaps the revised models will remove the unexplained free energy of nontransferable models and allow reaction schemes to fit data with one set of rate constants, over a range of conditions, so they become transferable and more useful in dealing with physiological situations.

Transferable models of this type would be helpful in practical applications. Biochemical reactions are usually studied in idealized laboratory conditions quite different from the situations (e.g., salt concentrations) present in biological cells. Models are needed that calculate rates correctly in conditions different from those in the laboratory. Big data analysis will find it difficult to remove errors introduced by rate constants that are very different from those assumed.

Models of semiconductor devices do quite well over a range of conditions. They are designed to satisfy Maxwell equations. Perhaps chemical reaction schemes can do as well once they are modified to conserve current and be consistent with the Maxwell equations.

Models that assume constant dielectric constants and permittivity hide the significance of conservation of current. Models with constant permittivity automatically satisfy conservation of current because they include a material displacement current that does not describe the displacement current in real materials, even approximately (over a range of conditions and times) as described above.

**Miracle of Maxwell's equations.** The miracle is that $\varepsilon_0 \partial \mathbf{E}/\partial t$ (vacuum displacement current) always assumes the value needed to guarantee conservation of current everywhere, at every time, in all conditions, with a single constant $\varepsilon_0$.

# Appendix: Misleading Formulations

It is tempting to oversimplify Maxwell's version of Ampere's law as

$$\mathbf{curl}\left(\mathbf{B}/\mu_0\right) = \mathbf{J}_{\text{transport}} + \varepsilon_r \varepsilon_0 \frac{\partial \mathbf{E}}{\partial t} \tag{A-1}$$

yielding the conservation law of ideal materials

$$\mathbf{div}\left(\mathbf{J}_{\text{transport}} + \varepsilon_r \varepsilon_0 \frac{\partial \mathbf{E}}{\partial t}\right) = 0 \tag{A-2}$$

Statements like A-2 tend to obscure the universal nature of conservation of current and so are inadvertently misleading.

'Everyone' knows that $\varepsilon_r$ varies from ~79 to 2 in water and is never a constant. So 'everyone' thinks that equations involving $\varepsilon_r$ are convenient over-simplifications and not universal laws at all. So 'everyone' thinks of conservation of current (A-2) in this light, as a fib, a white lie, hardly a universal law. But conservation of current can be written another way (see eq. 3) showing that it is always true. That is what this paper is about.

# References

Eisenberg, B., Yunkyong Hyon, and Chun Liu. 2010. 'Energy Variational Analysis EnVarA of Ions in Water and Channels: Field Theory for Primitive Models of Complex Ionic Fluids', *Journal of Chemical Physics*, 133: 104104

Eisenberg, Bob. 2006. 'The value of Einstein's mistakes. *Letter to the Editor:* "Einstein should be allowed his mistakes …" ', *Physics Today*, 59: 12.

———. 2014. 'Shouldn't we make biochemistry an exact science?', *ASBMB Today*, 13: 36-38.

Eisenberg, Robert S. 2016. 'Mass Action and Conservation of Current', *Hungarian Journal of Industry and Chemistry*, 44: 1-28 Posted on arXiv.org with paper ID arXiv:1502.07251.

Ferry, DK. 1980. 'Long-time tail of the autocorrelation function for electron drift in high electric fields in silicon', *Physical Review Letters*, 45: 758.

Feynman, R.P., R.B. Leighton, and M. Sands. 1963. *The Feynman: Lectures on Physics, Mainly Electromagnetism and Matter* (Addison-Wesley Publishing Co., also at http://www.feynmanlectures.caltech.edu/II_toc.html: New York).

Forster, Johannes. 2013. 'Mathematical Modeling of Complex Fluids', Master's, University of Wurzburg.

Gabrielse, Gerald, and David Hanneke. 2006. 'Precision pins down the electron's magnetism', *CERN Courier: International Journal of High-Energy Physics*.

Horng, Tzyy-Leng, Tai-Chia Lin, Chun Liu, and Bob Eisenberg. 2012. 'PNP Equations with Steric Effects: A Model of Ion Flow through Channels', *The Journal of Physical Chemistry B*, 116: 11422-41.

Jackson, J.D. 1999. *Classical Electrodynamics, Third Edition* (Wiley: New York).

Jimenez-Morales, David, Jie Liang, and Bob Eisenberg. 2012. 'Ionizable side chains at catalytic active sites of enzymes', *European Biophysics Journal*, 41: 449-60.

Ryham, R., Chun Liu, and Z. Q. Wang. 2006. 'On electro-kinetic fluids: One dimensional configurations', *Discrete and Continuous Dynamical Systems-Series B*, 6: 357-71.

Ryham, Rolf Josef. 2006. 'An Energetic Variational Approach to Mathematical Moldeling of Charged Fluids, Charge Phases, Simulation and Well Posedness, Ph.D. Thesis', Ph.D., The Pennsylvania State University.

Wang, Y., C. Liu, and Z. Tan. 2016. 'A Generalized Poisson--Nernst--Planck--Navier--Stokes Model on the Fluid with the Crowded Charged Particles: Derivation and Its Well-Posedness', *SIAM Journal on Mathematical Analysis*: 3191-235.

Wu, Hao, Tai-Chia Lin, and Chun Liu. 2014a. 'Diffusion Limit of Kinetic Equations for Multiple Species Charged Particles', *Archive for Rational Mechanics and Analysis*: 1-23.

———. 2014b. 'On transport of ionic solutions: from kinetic laws to continuum descriptions', *available on http://arxiv.org/ as 1306.3053v2*.

———. 2015. 'Diffusion Limit of Kinetic Equations for Multiple Species Charged Particles', *Archive for Rational Mechanics and Analysis*, 215: 419-41.

Xu, Shixin, Ping Sheng, and Chun Liu. 2014. 'An energetic variational approach to ion transport', *Communications in Mathematical Sciences*, 12: 779–89 Available on arXiv as http://arxiv.org/abs/1408.4114.